\documentclass[aip,pop,reprint]{revtex4-2}

\usepackage{amssymb}
\usepackage{graphicx}
\usepackage{bm}

\begin{document}


\title{A \textquotedblleft Galactic Disk\textquotedblright-Model for Three-Dimensional Bernstein-Greene-Kruskal Modes in a Finite Magnetic Field}


\author{C. S. Ng}
\email[]{cng2@alaska.edu}
\homepage[]{https://sites.google.com/a/alaska.edu/chungsangng/}
\affiliation{Geophysical Institute, University of Alaska Fairbanks, Fairbanks, Alaska 99775, USA}


\date{\today}

\begin{abstract}
A new model, 
  inspired by the structure of galactic disks, 
  for three-dimensional Bernstein-Greene-Kruskal (BGK) modes 
  in a plasma with a uniform finite background magnetic field 
  is presented.
These modes are exact nonlinear solutions of the steady-state Vlasov equation, 
  with an electric potential and a
  magnetic potential perturbation
  localized in all three spatial dimensions 
  that satisfies the Poisson equation,
  and the Amp\`{e}re Law,
  self-consistently.
The existence of solutions is shown analytically in the limit
  of small electric and magnetic field perturbations 
  associated with the disk species,
  and numerically 
  using an iterative method that
  converges up to moderately strong field perturbations.

\end{abstract}


\maketitle


\section{Introduction}
A fundamental problem in plasma physics is whether small
  kinetic scale structures can form in high temperature
  collisionless plasmas.
One possible solution for such structures was obtained by
  Bernstein, Greene, and Kruskal, known as the BGK mode
  \cite{PhysRev.108.546}.
The original BGK mode is a one-dimensional (1D)
  steady-state solution of the Vlasov-Poisson system of equations,
  which means the solution depends on one
  Cartesian coordinate
  such that the electric potential
  for a localized solution tends to zero as
  this coordinate tends to $\pm \infty$.
There have been a large volume of research on BGK modes
  than we can cite here in a short paper.
A  more comprehensive discussion on relevant literature can be found in
  our recent paper 
  \cite{doi:10.1063/1.5126705},
  as well as a recent review paper 
  \cite{doi:10.1063/1.4976854}.
Here we restrict our discussion on three-dimensional (3D)
  BGK modes,
  which have a localized electric potential that tends to
  zero along any direction in a 3D spatial space.
Motivated by multi-dimensional features observed in space plasmas
  \cite{PhysRevLett.81.826},
  a theory of 3D BGK modes was developed in a plasma 
  within a uniform magnetic field with infinite field strength
  \cite{2002PhDT........10C, doi:10.1029/2001GL013385, PhysRevE.69.055401},
  which has been further developed recently
  \cite{doi:10.1063/5.0045296}.
The original theory has imposed a cylindrical symmetry,
  mainly for convenience,
  so the solution depends on two coordinates.
It is considered 3D since the electric potential is localized
  in the sense mentioned above.
In the other extreme in the parameter space,
  another 3D BGK mode was found in an 
  unmagnetized (zero magnetic field) plasma
  \cite{PhysRevLett.95.245004}. 
That mode is still considered 3D despite having spherical symmetry
  such that the solution depends only on the spherical radial coordinate,
  because the electric potential tends to zero along all radial directions
  within a 3D spatial space.
The physical mechanism for this mode is different from other
  solutions discussed so far,
  since it is not based on particle trapping.
Instead,
  it is due to the dependence on another conserved quantity,
  the angular momentum associated with the spherical symmetry,
  in addition to the energy in the particle distribution function.
For a magnetic field with a finite field strength,
  in between the above two extremes,
  no exact 3D BGK mode solutions are found so far.
However,
  a theory of 2D BGK modes under this condition was developed
   \cite{doi:10.1063/1.2186187}, 
   with the distribution function depending also on the
   canonical angular momentum,
   which was later developed further
   \cite{doi:10.1063/1.5126705},
   with the additional dependence on the
   canonical momentum along the direction of 
   the background magnetic field.
These solutions are considered 2D since the localized
  electric potential tends to zero along any directions
  on a 2D plane perpendicular to the magnetic field direction,
  despite depending only on the spatial cylindrical radial 
  coordinate after imposing the cylindrical symmetry.
The stability of such solutions has been studied 
  using large-scale Particle-in-Cell (PIC) simulations,
  in both 2D
  \cite{10.1063/5.0187853}
  and 3D
  \cite{3D_BGK_Paper_arxiv}.
Both stable and unstable solutions were found in these
  simulations,
  with the formation of interesting spiral structures 
  during an instability,
  resembling spiral galaxies. 

The kinetic theory of a plasma turns out to be
  formally very similar to that of galactic dynamics
  such that 
  ``many of the mathematical tools that have been developed to
  study stellar systems are borrowed from plasma physics''
  \cite{10.2307/j.ctvc778ff}.
In the linear regime,
  fundamental concepts like Landau damping 
  and Case-Van Kampen modes
  are also very relevant to waves and instabilities 
  within a stellar system
  \cite{Ng_2021}.
In the nonlinear regime,
  a stellar equilibrium,
  such as a galaxy,
  can be obtained theoretically using a mathematical
  framework similar to constructing a 3D BGK mode
  \cite{PhysRevLett.95.245004}. 
One major difference between the two systems
  is that the gravitational interaction between stars is always 
  attractive,
  while the electric force is repulsive between charges of the same sign.
Because of this difference,
  finding an equilibrium solution for a stellar system
  turns out to be much easier than for a plasma.

In the development of a theory of 3D BGK modes
  for a magnetized plasma with a finite field strength,
  the direction of the influence of ideas is reversed here.
Our new model is inspired by the existence of
  disk galaxies,
  one of the most magnificent structures 
  in the universe that can be easily observed.
In this ``Galactic Disk''-model,
  the BGK mode is supported by introducing
  a localized disk species that is also spinning
  (due to the dependence of the 
  canonical angular momentum in the 
  distribution function of the disk species),
  similar to a disk galaxy 
  (although without spiral arms due to the
  imposed cylindrical symmetry).
In the following,
  we first present the mathematical formulation of this
  model,
  showing analytically the existence of exact solutions
  of the Vlasov-Poisson-Amp\`{e}re system of equations,
  in the limit of weak electric and magnetic field
  perturbations associated with the disk.
We then present numerical solutions
  beyond this limit,
  using an iteration scheme that converges
  up to a case with moderately large
  electric and magnetic field
  perturbations.
Possible implications of the new models 
  are then discussed,
  with the interesting possibility that
  the existence of disk-like structures
  in a plasma might provide 
  insights to fundamental problems
  in galactic dynamics.
  
\section{A \textquotedblleft Galactic Disk\textquotedblright-Model}
We start the construction of 3D BGK modes by requiring the distribution function
  $f_s$ for species $s$ to be a time-steady solution satisfying
  the Vlasov (Collisionless Boltzmann) equations,
\begin{equation}
  \frac{\partial f_s}{\partial t} + \textbf{v}\cdot\frac{\partial f_s}{\partial \textbf{r}} 
  + \frac{q_s}{m_s}(\textbf{E}+\textbf{v}\times\textbf{B})\cdot
  \frac{\partial f_s}{\partial \textbf{v}} = 0 \; ,  \label{vlasov}
\end{equation}
  where $q_s$ and $m_s$ are the charge and mass of the $s$ particle,
  with $f_s$ assumed to be independent of time $t$ so that 
  $f_s = f_s(\textbf{r}, \textbf{v})$ 
  depends only on spatial and 
  velocity space independent variables 
  $\textbf{r}$ and $\textbf{v}$.
While generally we can have as many $s$ species as we want,
  we will for simplicity restrict our study here to $s = i$, $e$, $d$,
  for the three species, ions, electrons, and the ``disk'' particles.
In the numerical examples presented in this paper, 
  we will further choose ions to be protons,
  and the $d$ particles to be another population of electrons,
  distinguished from the main $e$ species.
Under this choice,
  we have $m_i = m_p$, 
  the proton mass,
  $m_d = m_e$, 
  the electron mass,
  and $q_i = e > 0$,
  $q_e = q_d = -e$
  with $e$ being the magnitude of the electron charge.
For general expressions, 
  it is convenient to define $M_s = m_s/m_e$,
  and $Z_s = q_s/e$.
The electric field 
  $\textbf{E} = -\nabla \psi$ 
  and the magnetic field 
  $\textbf{B} = \nabla \times \textbf{A}$ 
  in Eq.~(\ref{vlasov}),
  or the electric potential $\psi$ and
  the magnetic potential $\textbf{A}$,
  are related self-consistently 
  and nonlinearly with $f_s$
  through the Gauss Law (Poisson equation)
  and the Amp\`{e}re Law,
\begin{eqnarray}
   \nabla^2 \psi =
    -\frac{\rho_q}{\epsilon_0} & = &
    -\frac{1}{\epsilon_0}\sum_s q_s \int d^3 v f_s  \; , 
     \label{gauss} 
     \\
   \nabla \times \nabla \times \textbf{A} =
      \mu_0 \textbf{J} & = &
      \mu_0 \sum_s q_s \int d^3 v f_s \textbf{v} \; .
     \label{ampere}
\end{eqnarray}
We are looking for localized 3D BGK modes 
  in a uniform background magnetic field 
  $\textbf{B} = B_\infty \hat{z}$,
  and a uniform background plasma with Maxwellian electrons and ions
  with density 
  $n_{e0}$ and $n_{i0} = n_{e0}/Z_i$,
  and temperature
  $T_{e0}$ and $T_{i0}$ respectively.
Therefore,
  we impose boundary conditions for $\psi$,
  and the magnetic potential perturbation 
  $\textbf{a} = \textbf{A} - 0.5 B_\infty \rho \hat{\phi}$,
  that they tend to zero as 
  $|\textbf{r}| \rightarrow \infty$
  along any direction in a 3D spatial space.
Note that we are using a cylindrical coordinate system
  with coordinates 
  $\rho$, $\phi$, $z$ 
  and corresponding
  unit vectors 
  $\hat{\rho}$, $\hat{\phi}$, $\hat{z}$.
The origin of the coordinate system is chosen to be the 
  center of the mode structure.
To simplify expressions,
  from now on we will normalize our equations
  using the electron thermal velocity
  $v_e = (k_B T_{e0}/m_e)^{1/2}$ as the unit for $\textbf{v}$,
  with $k_B$ being the Boltzmann constant
  and $T_{e0}$ the background electron temperature,
  the electron Debye length
  $\lambda_D = v_e/\omega_{pe} = v_e (\epsilon_0 m_e/n_{e0})^{1/2}/e$
  as the unit for $\textbf{r}$,
  $n_{e0}e\lambda^2_D/\epsilon_0$ as the unit for $\psi$,
  $n_{e0}e\lambda_D/\epsilon_0 v_e$ as the unit for $\textbf{B}$,
  $n_{e0}/v^3_e$ as the unit for $f_e$ and $f_d$, 
  and $n_{i0}/v^3_e$ as the unit for $f_i$.

To look for time steady solutions of Eq.~(\ref{vlasov}) for $f_s$,
  it is well known that we can simply require $f_s$ to depend
  only on conserved quantities.
One obvious conserved quantity is the total (kinetic plus electrostatic potential)
  particle energy with a normalized form proportional to
  $w_s = v^2/2 + \zeta_s \psi$,
  with $\zeta_s = Z_s/M_s$.
To make use of another conserved quantity,
  we need to impose a cylindrical symmetry
  such that physical quantities do not depend on $\phi$.
With this symmetry,
  another conserved quantity is the $z$-component
  of the canonical angular momentum,
  with a normalized form proportional to
  $l_s = 2\rho(v_\phi + \zeta_s A_\phi)$,
  where the subscript $\phi$ indicates the $\phi$-component.
Note that in the theory of 2D BGK modes
   \cite{doi:10.1063/1.2186187, doi:10.1063/1.5126705},
  with solutions also independent of $z$,
  there is another conserved quantity,
  i.e., the $z$-component of the canonical  momentum,
  with a normalized form proportional to $v_z + \zeta_s A_z$.
For the 3D case,
  with one less conserved quantity,
  the task of finding a solution satisfying Eqs.~(\ref{gauss}) 
  and (\ref{ampere}) becomes much more difficult.
In the theory of 3D BGK modes in the limit of  
  an infinitely strong background magnetic field
  \cite{2002PhDT........10C, doi:10.1029/2001GL013385, PhysRevE.69.055401},
  another conserved quantity is simply the radial coordinate $\rho$,
  since charged particles move only along the background magnetic
  field lines under this assumption.
To obtain a solution without the drastic assumption of infinite magnetic field strength,
  it would be very helpful to make use of another conserved quantity.
Inspired by the dynamics of galaxies,
  a dynamical system closely related to the Vlasov system
  considered here,
  we will employ the non-classical integral of motion
  (or the third integral)
  that allows the existence of thin galactic disks 
  (see Section 4.5 of Ref.~\onlinecite{10.2307/j.ctvc778ff}).
In contrast with classical integrals of motions $w_s$ and $l_s$
  discussed above,
  non-classical integrals of motion cannot be expressed as 
  functions of coordinates of the phase space.
Instead,
  they are expressed as a restriction of particle motions.
We impose this particular non-classical integral of motion
  by introducing a disk species with a 
  normalized distribution
\begin{equation}
 f_d=\frac{h_d}{2\pi\tau_d}\delta\left(v_z\right)\delta\left(z\right)e^{-k_dl_d-w_d/\tau_d}
  \; ,  \label{f-disk}
\end{equation}
  where $w_d$ and $l_d$ are defined as above with $s = d$,
  $\tau_d = v^2_d/v^2_e$ with $v_d$ being the thermal velocity
  of the particles of the disk species,
  $k_d$ and $h_d$ are constants.
Moreover,
  for a localized disk with
  $f_d \rightarrow 0$ as $\rho \rightarrow \infty$,
  the asymptotic magnetic field strength has to satisfy
  $B_\infty > 2k_d\tau_d/\zeta_d>0$.
One can check that the form of $f_d$ in Eq.~(\ref{f-disk}) satisfies 
  the Vlasov equation, Eq.~(\ref{vlasov}), 
  if $\psi$ and $\textbf{A}$ have the cylindrical symmetry,
  as well as a reflection symmetry with respect to the $z=0$ plane,
  i.e.,
\begin{equation}
 \psi\left(\rho,-z\right)=\psi\left(\rho,z\right), \,
 A_\phi\left(\rho,-z\right)=A_\phi\left(\rho,z\right)
  \; ,  \label{r-sym}
\end{equation}
  with $\textbf{A} = A_\phi (\rho,z) \hat{\phi}$,
  such that on the $z = 0$ plane 
  $\textbf{E} = E_\rho (\rho) \hat{\rho}$
  and $\textbf{B} = B_z (\rho) \hat{z}$.
Therefore,
  all disk particles will stay on the $z = 0$ plane,
  because their velocity and acceleration vectors are
  also on the same plane.
The dependence on $w_d$ and $l_d$ can be more general than in Eq.~(\ref{f-disk}).
The present form is chosen for simplicity so that the moments from $f_d$ 
  can be integrated into simple analytic expressions.
The disk species is also the first attempt to use a non-classical integral of motion
  that is most easily incorporated into the form of $f_d$.
Results from the current work thus suggest the possibility to employ
  other non-classical integrals of motion as well,
  although doing so might be significantly more difficult.

The ion and electron distribution functions,
 $f_i$ and $f_e$ can also be chosen as rather 
 general functions of $w_s$ and $l_s$.
For simplicity, 
  we set them in this paper to be Boltzmann distributions
\begin{equation}
 f_{e} = e^{-w_e}/\left(2\pi\right)^{3/2}, \,
 f_{i} = e^{-w_i/\tau_i}/\left(2\pi\tau_i\right)^{3/2}
  \; ,  \label{fefi}
\end{equation}
  where $\tau_i = v^2_i/v^2_e$ with $v_i$ being the ion thermal velocity.
The normalized charge density $\rho_q$ from Eqs.~(\ref{f-disk}) and (\ref{fefi}) 
  is then
\begin{equation}
  \rho_q=e^{-\zeta_i\psi/\tau_i}-e^\psi+\sigma\delta\left(z\right)
  \; ,  \label{rho-q}
\end{equation}
where $\sigma$ is a surface charge density on the $z=0$ plane,
\begin{equation}
  \sigma(\rho)=Z_dh_de^{-2k_d\rho\left(\zeta_dA_\phi-k_d\tau_d\rho\right)-\zeta_d\psi/\tau_d}
  \; .  \label{sigma}
\end{equation}
Similarly,
  because the disk has a rotational flow in the $\phi$-direction,
  there is a normalized current density $\textbf{J} = \nabla\times\mathbf{B}= 
  -\xi(\rho)\delta(z) \hat{\phi}$,
  which  is totally a surface current density,
  with
\begin{equation}
  \xi(\rho) = 2\beta_d^2k_d\rho Z_dh_de^{-2k_d\rho\left(\zeta_dA_\phi-k_d\tau_d\rho\right)-\zeta_d\psi/\tau_d}
  \; ,  \label{xi}
\end{equation}
where $\beta_d = v_d/c$ with $c$ being the speed of light in vacuum.
With the charge density given by Eq.~(\ref{rho-q}),
  the normalized form of Eq.~(\ref{gauss}) can be written as
\begin{equation}
  \frac{\partial}{\rho\partial\rho}\left(\rho\frac{\partial\psi}{\partial\rho}\right)+
  \frac{\partial^2\psi}{\partial z^2}=e^\psi-e^{-\zeta_i\psi/\tau_i}
  \; ,  \label{psi-eq}
\end{equation}
  for $z >0$ with $\psi\left(\rho,0\right)=\psi\left(\rho,z\rightarrow0\right)$.
$\psi (\rho, z)$ for $z < 0$ is given by 
  the symmetry condition Eq.~(\ref{r-sym}).
The solution is subjected to boundary conditions
  $\psi\left(\rho\rightarrow\infty,z\right) \rightarrow 0$,
  $\psi\left(\rho,z\rightarrow\infty\right) \rightarrow 0$,
  as well as
\begin{equation}
  \frac{\partial\psi\left(\rho,0^+\right)}{\partial z}=-\sigma\left(\rho;\psi,a_\phi\right)/2, \;
  \frac{\partial\psi\left(0,z\right)}{\partial\rho}=0
  \; ,  \label{psi-bc}
\end{equation}
  where the semicolon notation in $\sigma$ is to indicate that it depends on $\psi$
  and $a_\phi$ also,
  with $a_\phi = A_\phi - B_\infty \rho /2$,.
Similarly, 
  the normalized form of Eq.~(\ref{ampere}) can be written as
\begin{equation}
  \frac{\partial^2a_\phi}{\partial\rho^2}+\frac{1}{\rho}\frac{{\partial a}_\phi}{\partial\rho}
  -\frac{a_\phi}{\rho^2}+\frac{\partial^2a_\phi}{\partial z^2}=0
  \; ,  \label{a-eq}
\end{equation}
for $z >0$ with $a_\phi\left(\rho,0\right)=a_\phi\left(\rho,z\rightarrow0\right)$
  and the symmetry condition Eq.~(\ref{r-sym}),
  subject to boundary conditions  
  $a_\phi\left(\rho\rightarrow\infty,z\right) \rightarrow 0$,
  $a_\phi\left(\rho,z\rightarrow\infty\right) \rightarrow 0$,
  as well as
\begin{equation}
  \frac{\partial a_\phi\left(\rho,0^+\right)}{\partial z}= \xi\left(\rho;\psi,a_\phi\right)/2, \;
  a_\phi\left(\rho\rightarrow 0,z\right) \rightarrow 0
  \; .  \label{a-bc}
\end{equation}
Eqs.~(\ref{psi-eq}) to (\ref{a-bc}) thus form a set of nonlinear
  partial differential equations for the two functions
  $\psi(\rho,z)$ and $a_\phi(\rho,z)$.
Due to the complicated structure,
  it is likely that solutions can only be found numerically.
While there can be many different numerical schemes for this problem,
  here we will only present a straightforward one by using Hankel transforms
\begin{eqnarray}
  \psi\left(\rho,z\right) & =& \int_{0}^{\infty}kdk\widetilde{\psi}\left(k,z\right)J_0\left(k\rho\right), \; 
  \nonumber
  \\
  a_\phi\left(\rho,z\right) & = & \int_{0}^{\infty}kdk{\widetilde{a}}_\phi\left(k,z\right)J_1\left(k\rho\right)
  \; ,  \label{hankel}
\end{eqnarray}
  where $J_0$ and $J_1$ are Bessel functions of zeroth and first order respectively.
Eqs.~(\ref{psi-eq}) and (\ref{psi-bc}) can then be written as
\begin{equation}
  \frac{\partial^2\widetilde{\psi}\left(k,z\right)}{\partial z^2}-
  \left(k^2+1+\frac{\zeta_i}{\tau_i}\right)\widetilde{\psi}\left(k,z\right)=-\gamma\left(k,z\right)
  \; ,  \label{psi-k-eq}
\end{equation}
  where $\gamma\left(k,z\right)=\int_{0}^{\infty}{\rho d\rho\Gamma(\rho,z)J_0\left(k\rho\right)}$,
  with $\Gamma\left(\rho,z\right)=\exp ({-\zeta}_i\psi/\tau_i) - \exp (\psi)+\left(1+\zeta_i/\tau_i\right)\psi$,
  subject to the Neumann boundary condition on the $z = 0$ plane
\begin{equation}
  \frac{\partial\widetilde{\psi}\left(k,0^+\right)}{\partial z}=-\frac{1}{2}\int_{0}^{\infty}
  {\rho d\rho\sigma\left(\rho;\psi,a_\phi\right)J_0\left(k\rho\right)}
  \; .  \label{psi-k-bc}
\end{equation} 
Since $\gamma\left(k,z \rightarrow \infty \right) \rightarrow 0$, 
  the solution $\widetilde{\psi} \propto \exp \left[ -\left(k^2+1+\zeta_i /\tau_i\right)^{1/2} z\right]$ will be
  chosen for Eq.~(\ref{psi-k-eq}) in the $z\rightarrow\infty$ limit,
  in order to satisfy the boundary condition 
  $\psi\left(\rho,z\rightarrow\infty\right) \rightarrow 0$.
Similarly,
  Eqs.~(\ref{a-eq}) and (\ref{a-bc}) become
  ${\widetilde{a}}_\phi\left(k,z\right)={\widetilde{a}}_\phi\left(k,0\right)\exp\left(-kz\right)$,
  with
\begin{equation}
   {\widetilde{a}}_\phi\left(k,0\right)=-\frac{1}{2k}\int_{0}^{\infty}
   {\rho d\rho\xi\left(\rho;\psi,a_\phi\right)J_1\left(k\rho\right)}
  \; .  \label{a-k-bc}
\end{equation} 
By employing Hankel transforms,
  we therefore reduce the partial differential operator in Eq.~(\ref{psi-eq})
  to an ordinary differential operator in Eq.~(\ref{psi-k-eq}),
  but pay the price of making it also an integral equation through
  the right-hand-sides of
  Eqs.~(\ref{psi-k-eq}) to (\ref{a-k-bc}).
There is also an advantage that all boundary conditions can 
  now be satisfied in a straightforward manner.
Moreover,
  this formulation allows us to write down the solution
  in the weak field limit in closed form.
From Eqs.~(\ref{sigma}) and (\ref{xi}) we see that in the limit of 
  $h_d \rightarrow 0$,
  both the surface charge density and surface current density tend
  to zero.
We should then expect $\psi\rightarrow\psi_0=0$,
  and $a_\phi\rightarrow a_{\phi0}=0$
  such that $A_\phi\rightarrow A_{\phi0}=B_\infty\rho/2$. 
The right-hand-sides of
  Eqs.~(\ref{psi-k-bc}) and (\ref{a-k-bc}) no longer depend on
  $\psi$ and $a_\phi$
  if we set $\sigma\left(\rho;\psi,a_\phi\right)\rightarrow\sigma\left(\rho;\psi_0,a_{\phi0}\right)
  =Z_dh_d \exp \left[ -k_d\left(\zeta_dB_\infty-2k_d\tau_d\right)\rho^2 \right]$,
  and  $\xi\left(\rho;\psi,a_\phi\right)\rightarrow\xi\left(\rho;\psi_0,a_{\phi0}\right)
  =2\beta_d^2k_d\rho Z_dh_d \exp \left[ -k_d\left(\zeta_dB_\infty-2k_d\tau_d\right)\rho^2 \right]$.
In the same way,
  the right-hand-side of Eq.~(\ref{psi-k-eq}) becomes zero because
  $\Gamma \rightarrow O(\psi_0^2) \rightarrow 0$.
The solutions for $\psi(\rho,z)$ and $a_\phi(\rho,z)$ in the limit of $h_d \rightarrow 0$
  can then be solved analytically as integrals of the forms of
\begin{eqnarray}
   \psi\left(\rho,z\right) & = &
   \frac{Z_dh_d}{4k_d\left(\zeta_dB_\infty-2k_d\tau_d\right)}
   \int_{0}^{\infty}\frac{kdkJ_0\left(k\rho\right)}{\sqrt{k^2+1+\frac{\zeta_i}{\tau_i}}}
   \nonumber
   \\
   && e^{\frac{-k^2}{4k_d\left(\zeta_dB_\infty-2k_d\tau_d\right)}-|z|\sqrt{k^2+1
   +\frac{\zeta_i}{\tau_i}}}  \; , 
     \label{psi-hd0} 
     \\
   a_\phi\left(\rho,z\right) & = &
   \frac{-\beta_d^2k_dZ_dh_d}{\left[2k_d\left(\zeta_dB_\infty-2k_d\tau_d\right)\right]^2}
   \int_{0}^{\infty}kdkJ_1\left(k\rho\right)
    \nonumber
   \\
   &&
   e^{\frac{-k^2}{4k_d\left(\zeta_dB_\infty-2k_d\tau_d\right)}-k|z|} \; .
     \label{a-hd0}
\end{eqnarray}
It is straightforward to verify that the integrals in Eqs.~(\ref{psi-hd0})
  and (\ref{a-hd0}) are well behaved such that 
  3D BGK solutions do exist in the weak field limit
  $h_d \rightarrow 0$.
To show the existence of solutions for a finite $h_d$,
  one can try putting solutions from Eqs.~(\ref{psi-hd0})
  and (\ref{a-hd0}) back into the right-hand-sides of
  Eqs.~(\ref{psi-k-eq}) to (\ref{a-k-bc})
  and then solve Eq.~(\ref{psi-k-eq}) numerically
  in a straightforward manner,
  because the equation becomes an linear ODE for a
  new $\widetilde{\psi}$ after using the old $\psi$ and $a_\phi$ on 
  the right-hand-side.
The new solutions for $\psi(\rho,z)$ and $a_\phi(\rho,z)$ can then
  be put back to the iteration scheme to improve the solutions until they 
  converge.
In the following,
  we present one example to show that this iteration
  scheme does result in converged solutions up to a rather large $h_d$.

\begin{figure}
  \centerline{\includegraphics[width=\columnwidth]{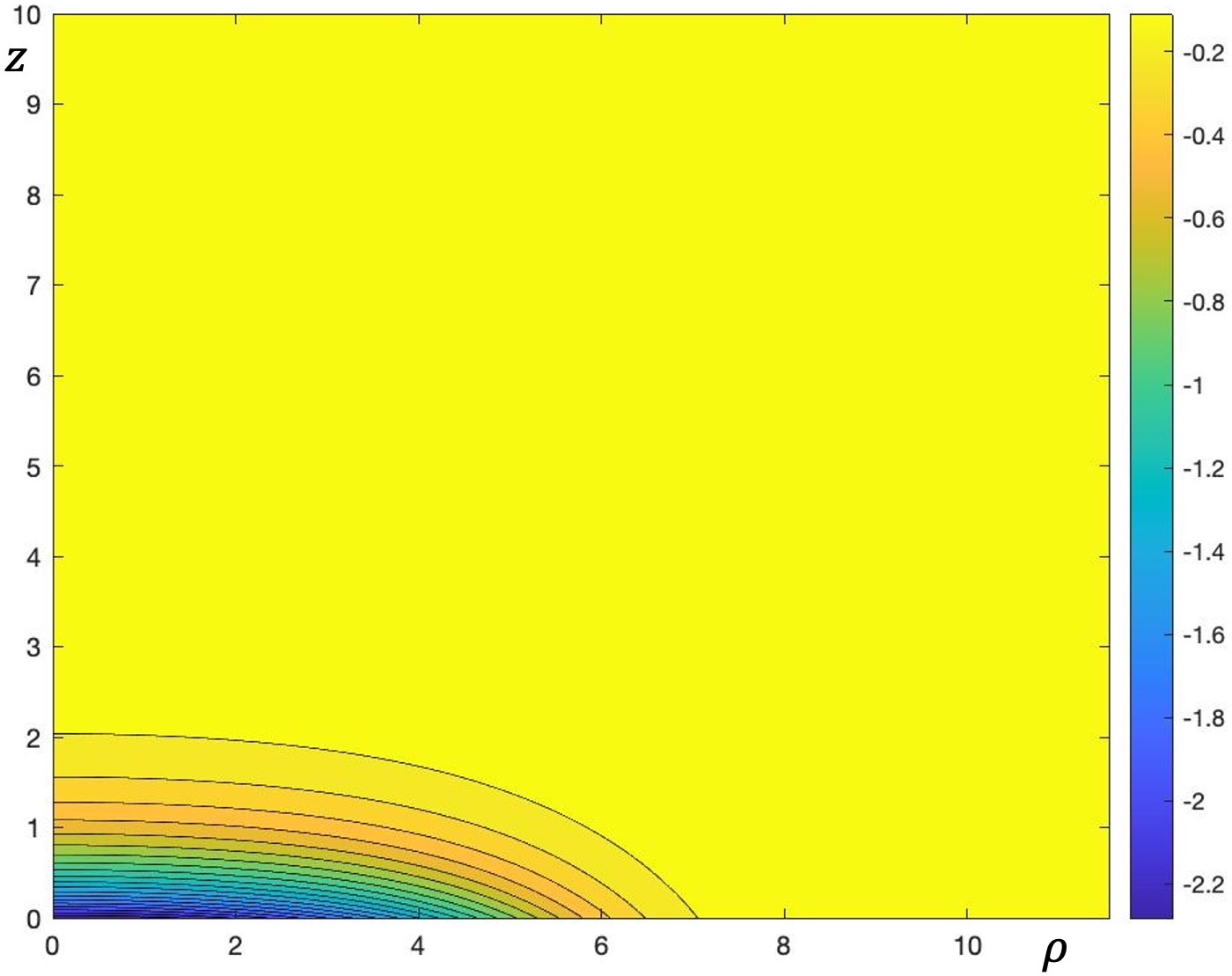}}
  \caption{Color coded contour plot 
  of the electric potential $\psi (\rho,z)$ 
  solved from a run with 
  $B_\infty = 1$,
  $\zeta_i=\tau_i=1/1836$,
  $\zeta_d=Z_d=-1$,
  $\beta_d^2 = 0.01$,
  $\tau_d=1$,
  $k_d=-0.3$,
  and $h_d = 79$.
  \label{psi-plot}}
\end{figure}
  
\begin{figure}
  \centerline{\includegraphics[width=\columnwidth]{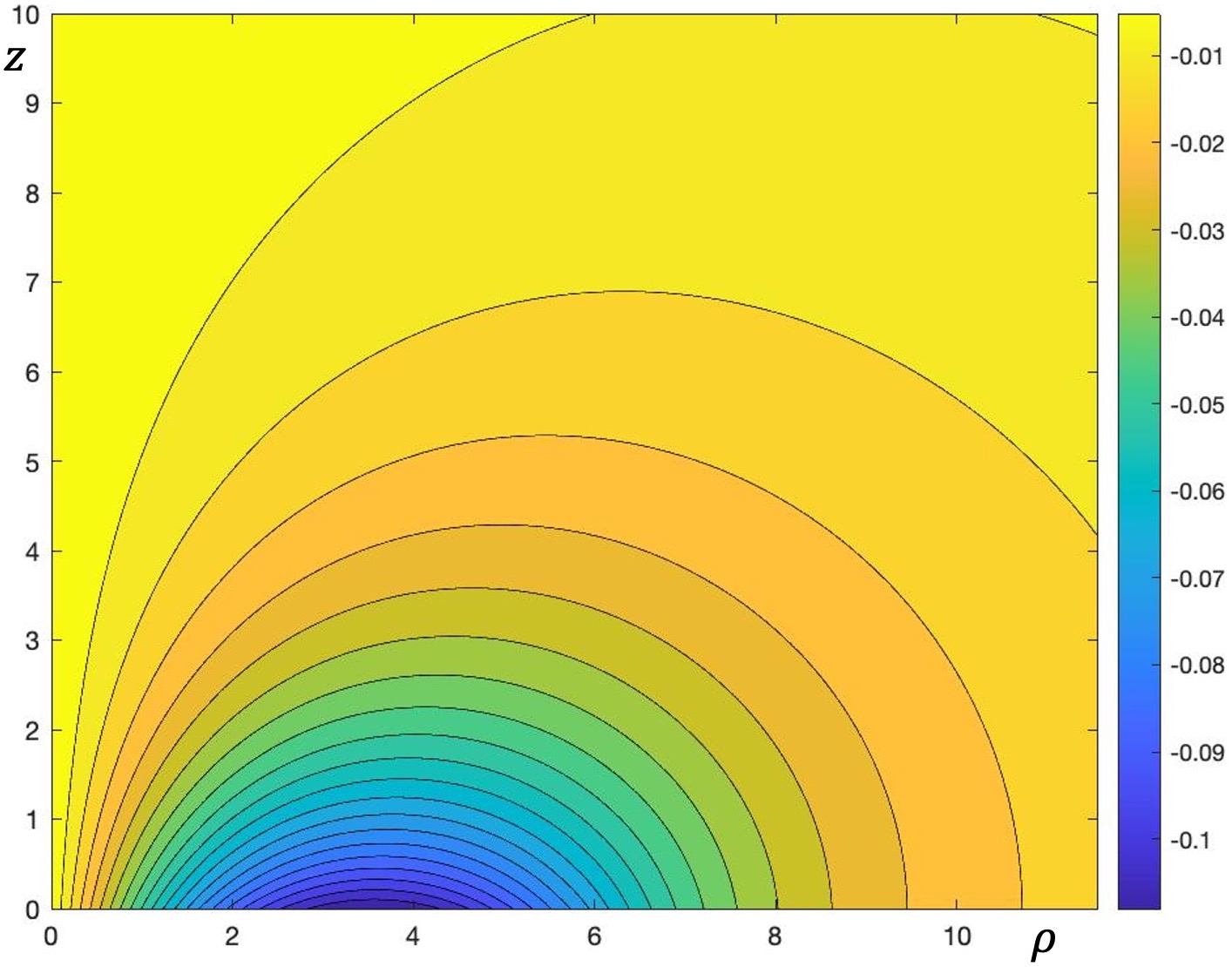}}
  \caption{Color coded contour plot 
  of the magnetic potential perturbation $a_\phi (\rho,z)$ 
  solved from the same run for Fig.~\ref{psi-plot}.
  \label{aphi-plot}}
\end{figure}

\section{Numerical Solutions}
Numerical solutions $\psi(\rho,z)$ and $a_\phi(\rho,z)$ 
  for Eqs.~(\ref{hankel}) to (\ref{a-k-bc}) are calculated on a uniform grid 
  over a domain $\rho = (0,\rho_{max})$, 
  and $z = (0,z_{max})$.
The domain is chosen such that $\psi(\rho,z)$ and $a_\phi(\rho,z)$
  are expected to be small for $\rho > \rho_{max}$,
  and $z > z_{max}$.
Direct and inverse Hankel transforms are performed using 
  numerical functions for $J_0$ and $J_1$,
  and integrated using an extended trapezoidal rule 
  \cite{press1996numerical}.
Because the main goal of this study is to show the existence of solutions,
  we have chosen to use simple straightforward numerical methods
  so that results can be easily verified.
Therefore,
  we have not employed a fast Hankel transform method,
  which would require other more complicated numerical implementation
  such as a nonuniform grid.
However,
  such a method to increase numerical efficiency can certainly be
  included in the future development of the code.
We have confirmed the accuracy of our method in 
  performing Hankel transforms,
  by comparing an analytical test function with itself
  after a pair of direct and inverse transforms.
Eq.~(\ref{psi-k-eq}),
  with the boundary condition Eq.~(\ref{psi-k-bc}),
   is solved numerically as a linear ODE,
   with the right-hand-sides treated as known terms,
   using a second order finite difference discretization solved by
   a tridiagonal matrix method 
   \cite{jaluria1986computational},
   to make sure that $\widetilde{\psi}$ is on a branch that 
   is exponentially small for large $z$.

As outlined briefly above,
  the iteration scheme starts with setting
  $\psi = \psi_i$ and $a_\phi = a_{\phi i}$
  on the right-hand-sides of
  Eqs.~(\ref{psi-k-eq}) to (\ref{a-k-bc}).
New solutions $\psi = \psi_f$ and $a_\phi = a_{\phi f}$
  are then solved by the method described above
  on the uniform grid.
The convergency of solutions is tested by calculating 
  a deviation parameter
\begin{equation}
   d = \left[ \frac{ 2 \left< \left(\psi_f - \psi_i \right)^2 \right>}
   { \left< \psi_f^2 \right>  +  \left< \psi_i^2 \right>}
   \right]^{1/2}
   +  \left[ \frac{ 2 \left< \left(a_{\phi f} - a_{\phi i} \right)^2 \right>}
   { \left< a_{\phi f}^2 \right>  +  \left< a_{\phi i}^2 \right>}
   \right]^{1/2}
  \; ,  \label{d}
\end{equation} 
  as a measure of the fractional difference between the old and new solutions,
  where the notation $\left< \dots \right>$ indicates a sum over
  all grid points of the quantity within the angle brackets. 
Obviously,
  the method converges if $d = 0$.
For a finite $d$,
  one then set $\psi_i = \psi_f$ and $a_{\phi i}= a_{\phi f}$
  and repeat the process in the hope of getting a smaller $d$.
In practice,
  solutions are considered converged if $d$ is smaller than 
  a certain tolerance.
To start the first iteration,
  the easiest choice is $\psi_i =a_{\phi i}= 0$,
  which is equivalent to solving for the
  zeroth order 
  (or $h_d \rightarrow 0$) solutions given by 
  Eqs.~(\ref{psi-hd0}) and (\ref{a-hd0}).
For a larger $h_d$,
  a faster choice is to use converged
  solutions $\psi$ and $a_\phi$ for a slightly
  smaller value of $h_d$,
  if they are available.

\begin{figure}
  \centerline{\includegraphics[width=\columnwidth]{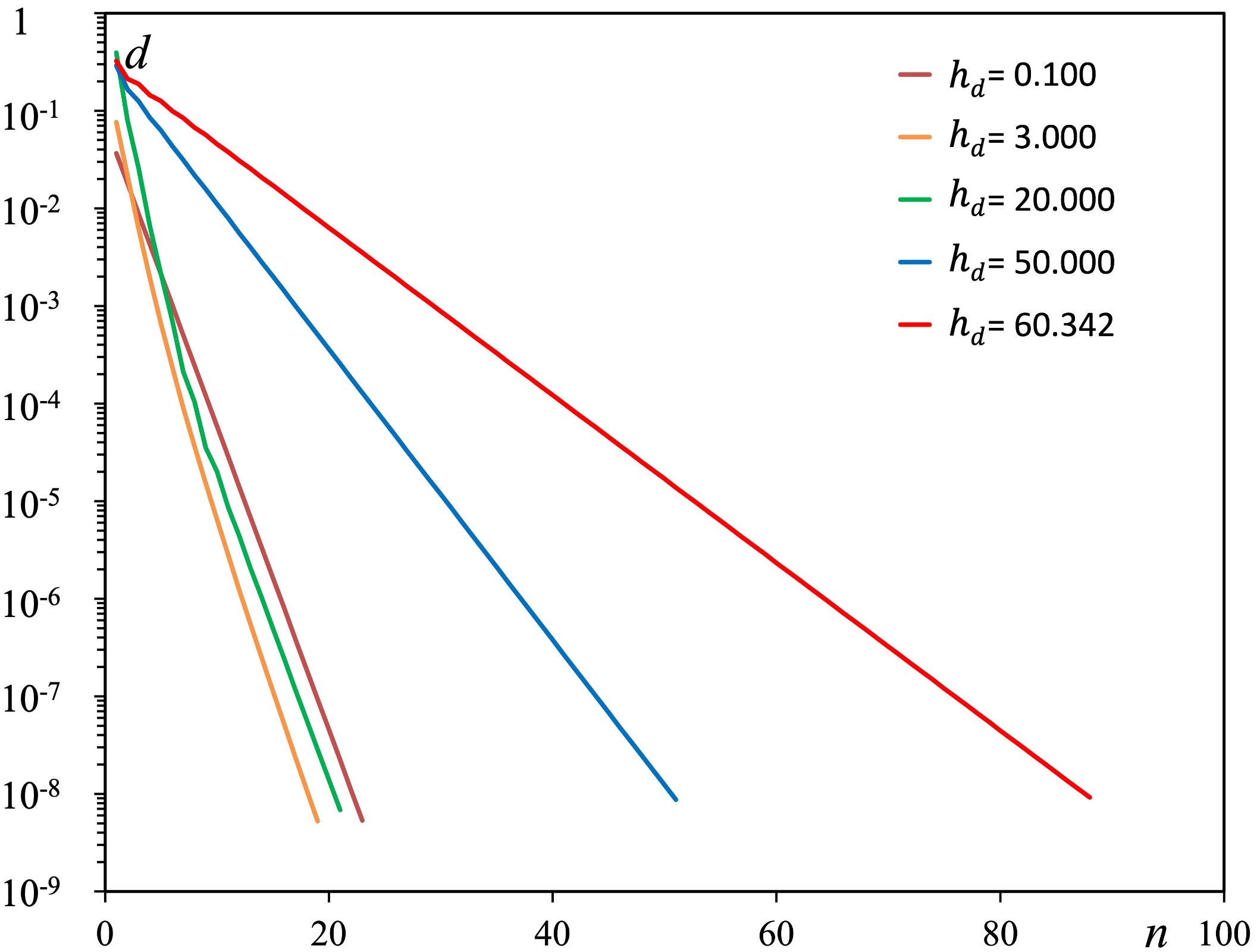}}
  \caption{The deviation parameter $d$ as a function
  of the number of iteration $n$ for some runs with 
  different $h_d$,
  but with other parameters the same 
  as in the run for Figs.~\ref{psi-plot} and \ref{aphi-plot}.
  \label{d_vs_n}}
\end{figure}

\begin{figure}
  \centerline{\includegraphics[width=\columnwidth]{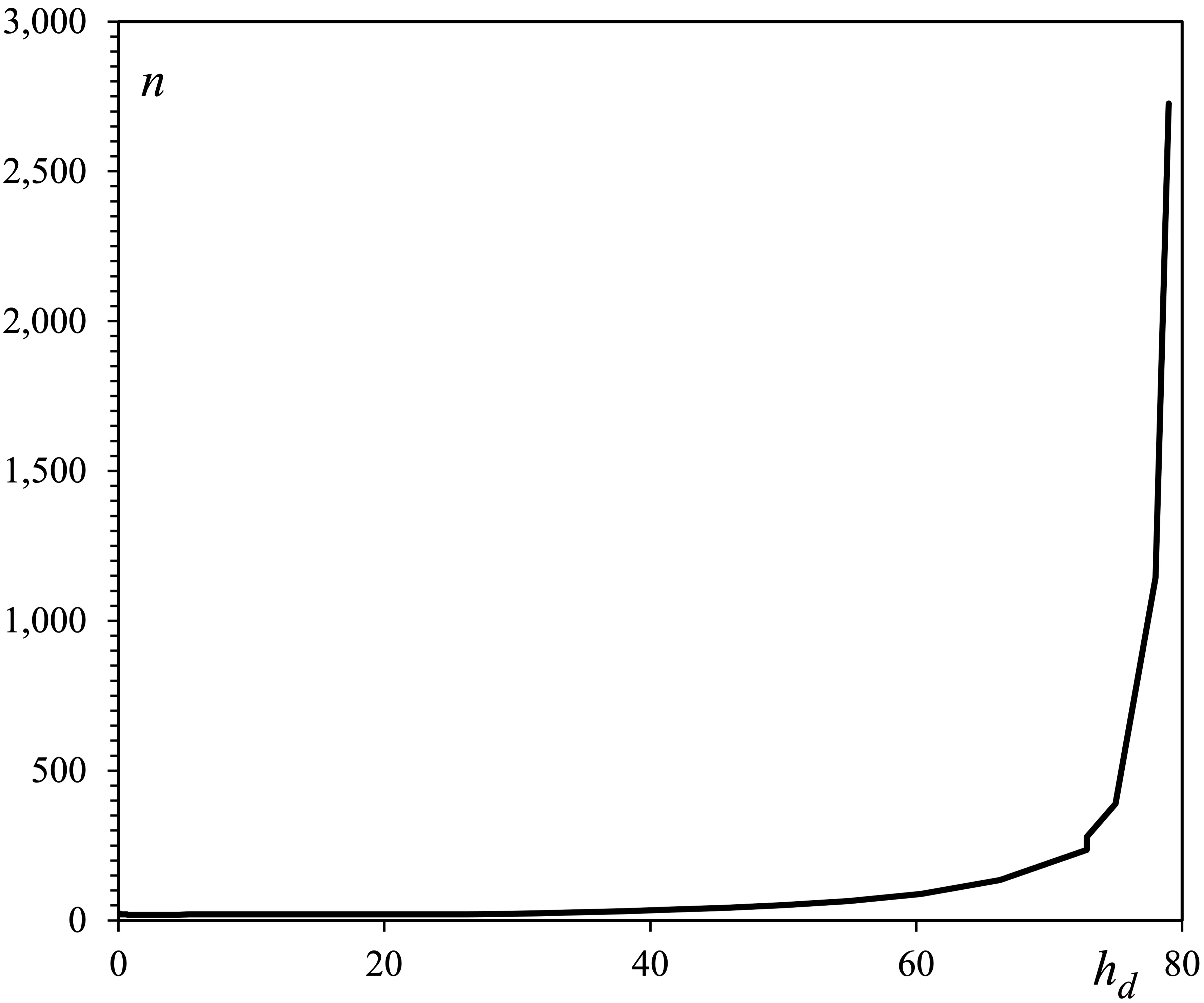}}
  \caption{The number of iterations $n$ it takes for $d < 10^{-8}$  
  as a function of $h_d$,
  using parameters the same as in Fig.~\ref{d_vs_n}.
  \label{n_vs_hd}}
\end{figure}
  
In the following,
  we present a specific case showing that this 
  iteration scheme is convergent up to a 
  moderately large $h_d$ value.
The parameters for this case are
  $B_\infty = 1$,
  $\zeta_i=\tau_i=1/1836$,
  $\zeta_d=Z_d=-1$,
  $\beta_d^2 = 0.01$,
  $\tau_d=1$,
  and $k_d=-0.3$.
This represents a proton plasma with an electron disk species,
  and with all species having the same background temperature.
Obviously,
  this set of parameters can take many different values.
However,
  just this one case is enough
  for our purpose to show the existence of solutions.
Fig.~\ref{psi-plot} shows a color coded contour plot 
  of the solution $\psi (\rho,z)$ for a run with these
  parameters and $h_d = 79$,
  over a $1024^2$ grid with
  $\rho_{max} = 11.547$,
  and $z_{max} = 10$.
The contour plot of $a_\phi (\rho,z)$ for this case is
  shown in Fig.~\ref{aphi-plot}.
This run stops at the iteration when the $d$ value is just below
  the tolerance of $10^{-8}$.
Many more runs with these parameters but with smaller $h_d$
  were performed.
Fig.~\ref{d_vs_n} shows how $d$ changes with the number of 
  iteration $n$ for some of these runs.
We see in Fig.~\ref{d_vs_n} that $d$ decreases exponentially 
  with $n$ as $d \rightarrow 0$.
However,
  the convergence becomes slower as $h_d$ increases.
The $h_d = 0.1$ case in Fig.~\ref{d_vs_n} uses
  the iteration method as described above by setting
  $\psi_i = \psi_f$ and $a_{\phi i}= a_{\phi f}$ in the
  next iteration.
It was found later that the convergence can get faster
  by setting the new $\psi_i$ and $a_{\phi i}$
  as $(\psi_i + \psi_f)/2$ and  $(a_{\phi i} + a_{\phi f})/2$
  instead.
This method was used for $h_d = 3$,
  which is why $d$ for this case
  decreases faster than the $h_d = 0.1$ case.
Obviously,
  this method does not change the determination of whether
  the solution is converged,
  as long as $d \rightarrow 0$.
The reason why the new method has a better convergence 
  can be traced to the fact that each iteration has a tendency
  of overshooting the ``true'' solution so that an average
  between the old and new solutions can result in a trial
  closer to the ``true'' solution.

\begin{figure}
  \centerline{\includegraphics[width=\columnwidth]{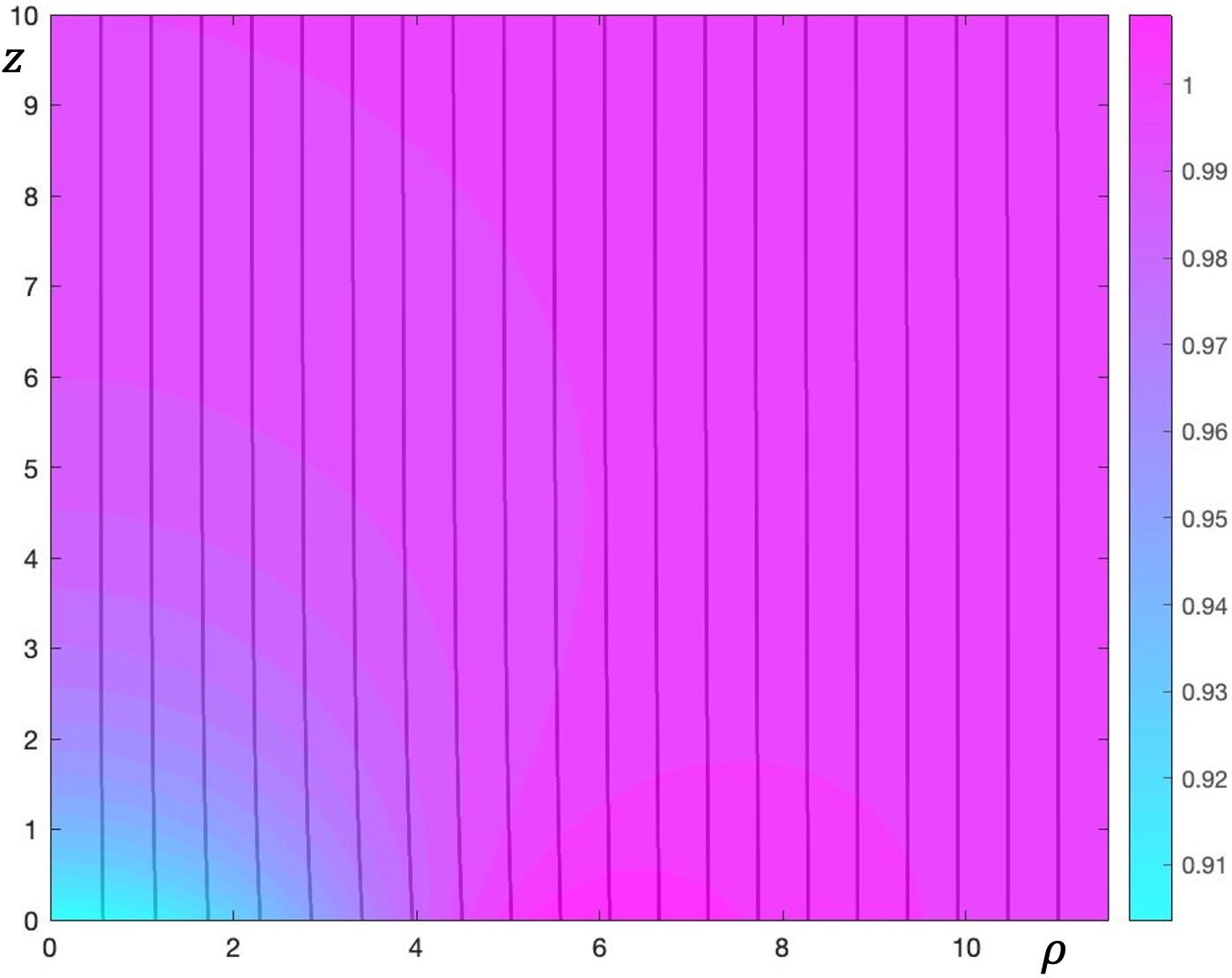}}
  \caption{Color coded map for 
  the total magnetic field strength $B$,
  overlaying with magnetic field lines (dark curves),
  from the same run for Figs.~\ref{psi-plot} and \ref{aphi-plot}.
  \label{B_total}}
\end{figure}
    
Fig.~\ref{d_vs_n} does not include the $h_d =79$ case shown
  in Figs.~\ref{psi-plot} and \ref{aphi-plot}
  because that case converges very slowly and it takes
  $n = 2726$ iterations for $d$ to decrease below $10^{-8}$.
To show this trend,
  we plot the number of iterations $n$ it takes for $d < 10^{-8}$  
  as a function of $h_d$ in Fig.~\ref{n_vs_hd}.
Due to the drastic increase of the number of iterations for 
  solutions to converge as shown in Fig.~\ref{n_vs_hd},
  we have not been able to
  obtain converged solutions with $h_d > 79$.
Currently it is not known whether this is an absolute
  mathematical convergence limit,
  or if it can be extended by employing a different
  numerical method to obtain solutions.
For our purpose,
  the existence of solutions is shown if converged
  solutions can be found for any finite $h_d$ values.
The largest $h_d$ value of 79 in fact is corresponding
  to moderately large electric and magnetic field perturbations.
Because we are using normalized quantities,
  the fact that the $\psi$ solution shown in 
  Fig.~\ref{psi-plot} has a
  magnitude larger than 2 with spatial scale of the 
  order of unity shows that the electric field is
  also of the order of unity,
  in the normalized unit.
For $a_\phi$,
  the total magnetic field is the curl of it plus the background
  uniform magnetic field with a strength $B_\infty = 1$.
Fig.~\ref{B_total} shows the color coded map for 
  the total magnetic field strength $B$,
  overlaying with magnetic field lines (dark curves).
We see that $B$ has a decrease of about 10 \% 
  of $B_\infty$ around the origin,
  the center of the structure.
Correspondingly,
  magnetic field lines have slight bends
  around the same region,
  although overall it looks very similar to a
  uniform magnetic field.
The magnetic structure of such a BGK mode
  is diamagnetic,
  independent of the charge of the
  disk species,
  different from the sign of $\psi$.
We confirm this with some other runs using ions
  as the disk species (not shown in this paper).
  
Finally,
  we remark that numerical solutions presented above
  have been verified in two other ways.
Firstly,
  we have run some cases with selected $h_d$ values,
  including the $h_d =79$ case,
  on a $2048^2$ grid with both $\rho_{max}$ and 
  $z_{max}$ increased by a factor of 1.5.
Effectively the linear resolution for the new runs 
  is also a factor of 1.5 higher than the old runs.
It is confirm that both the new and the old run 
  with the same $h_d$ have essentially the same
  $\psi$ and $a_\phi$ over the old domain,
  and with these fields extend to decreasingly
  small values in the new domain outside
  the old domain.
Using the higher resolution and larger domain
  also does not change the apparent 
  convergence limit of $h_d =79$.
We therefore confirm that the size of the domain
  and the resolution used in the old runs are
  adequate enough to produce accurate solutions.
Secondly,
  we substitute the final converged $\psi$ and $a_\phi$ solutions
  for cases with selected $h_d$ values,
  again including the $h_d =79$ case,
  into the set of Eqs.~(\ref{psi-eq}) to (\ref{a-bc}),
  with derivatives evaluated using second order finite differencing.
We confirm that the sum of all terms for each of these equations
  is a small fraction of the sum of the absolute value of these terms.
This provides a test of the solutions independent of the use of
  Hankel transforms,
  although we are not solving solutions using a separate
  numerical scheme based on a fully finite difference method.
 
\section{Discussion and Conclusion}
In this paper,
  we have presented a new ``Galactic Disk''-model for constructing 3D
  BGK mode solutions within a finite magnetic field.
The main ingredient of this model that enables the existence of solutions
  is a rotating thin disk species,
  inspired by the existence of thin disk galaxies,
  with the recognition of the similarities between galactic dynamics
  and the physics of a kinetic plasma.
Now that the existence of similar disk structures 
   in a plasma is shown theoretically,
   it is of fundamental interest to find out whether such
   structures can exist in nature,
   or be constructed in a laboratory experiment.
This is important  
   not only to small-scale kinetic physics of plasmas,
   but also to the large-scale galactic dynamics.
If experimental studies of disk structures in a plasma are possible,
   it can in turn provide insights to the dynamics of disk galaxies.
There are still fundamental unsolved problems,
   such as the formation of spiral arm structures,
   partially due to the enormous time scale in galactic dynamics,
   making it difficult to observe how galaxies evolve in time.
 Studies of the dynamics of plasmas obviously do not suffer 
   from the same problem.
 
This time-steady equilibrium of the disk species is enabled by a 
  non-classical integral of motion,
  and maintained by a reflection symmetry.
Therefore,
  this new model can potentially be a first step in
  trying to construct 3D solutions using other 
  non-classical integrals of motion,
  which might allow more regular forms of solutions
  than the discontinuous delta-function form in
  the current model.

We have shown the existence of solutions for this model
  by presenting numerical solutions for one example
  with an electron disk species.
The numerical method in solving these solutions is chosen
  because of simplicity rather than efficiency.
Solutions are checked with an increase of the linear resolution
  and the domain size,
  as well as a direct substitution into equations using
  finite difference for derivatives.
Therefore,
  the existence of solutions have been carefully verified,
  with solutions converged up to a case with
  moderately strong electric and magnetic field 
  perturbations.
We have not attempted to develop the numerical method further
  to solve for solutions more efficiently,
  or to extend the convergence limit of the iteration scheme.
However,
  this should be an interesting direction for future research,
  to see whether solutions exist for even larger $h_d$
  such that the magnetic field reverses direction at the center.
The magnetic configuration for a self-consistent
  localized electron-scale equilibrium 
  with such a magnetic field reversal 
  would be like a spheromak.   
  
The physical mechanism of this 3D BGK mode for a finite magnetic field
  is very different from the 3D BGK mode for a zero magnetic field 
  \cite{PhysRevLett.95.245004}, 
  or the 3D BGK mode for an infinite magnetic field 
  \cite{2002PhDT........10C, doi:10.1029/2001GL013385, PhysRevE.69.055401}.
It is quite unlikely that our solutions can tend to these solutions
  in the zero or infinite field strength limits.
Therefore,
  this shows that there can be multiple physical mechanisms
  for the existence of different forms of 3D BGK modes
  that do not tend to each other continuously.
The physical mechanism is more similar to the ``magnetic hole'' type
  2D BGK mode solutions 
  \cite{doi:10.1063/1.2186187, doi:10.1063/1.5126705, 3D_BGK_Paper_arxiv},
  except now the added electron vortex is only on a single plane
  while it is invariant along the whole $z$ direction for the 
  2D case.

Just as in many other theories of time-steady equilibrium,
  the existence of solutions in this model does not
  answer important questions regarding how such structures
  can form,
  and whether they are dynamically stable.
Because these questions involve the time evolution of kinetic plasmas,
  they most likely can only be answered definitely by 
  direct numerical simulations.
Studies of the stability problem using PIC simulations for the 
  2D BGK modes have been conducted
  \cite{doi:10.1063/1.4723590, 10.1063/5.0187853, 3D_BGK_Paper_arxiv}, 
  up to relatively high resolutions in both 2D and 3D runs.
Both stable and unstable solutions have been found,
  depending on parameters.
Generally a solution becomes unstable when the background
  magnetic field strength is smaller than a certain value,
  while keeping all other parameters fixed.
It was also found that when a solution becomes unstable,
  the electron density and electric field perturbation form
  a spiral structure resembling a spiral galaxy,
  when looking at a 2D plane perpendicular to the background
  magnetic field.
The stability study for this ``Galactic Disk''-model is certainly 
  an interesting extension of this effort.
Even if these disk structures turn out to be unstable,
  there can still be important transient phenomena,
  such the formation of spiral structures similar to what was
  observed in simulations of 2D BGK modes,
  since the mathematical form of the disk species distribution is
  the same as a 2D BGK mode.
Moreover,
  it can still be very interesting to find out
  what a 3D BGK mode would relax to,
  after an instability takes its course.

\section*{Data availability statement} 
Computing codes and data producing results for this paper are available at
\url{https://github.com/chungsangng/gd}.

\begin{acknowledgments}
This work is supported by a National Science Foundation grant PHY-2010617.
\end{acknowledgments}

\providecommand{\noopsort}[1]{}\providecommand{\singleletter}[1]{#1}%

\end{document}